\documentclass[twoside]{article}

\usepackage{blindtext} 

\usepackage[sc]{mathpazo} 
\usepackage[T1]{fontenc} 
\usepackage{microtype} 

\usepackage[english]{babel} 

\usepackage[hmarginratio=1:1,top=32mm,columnsep=20pt]{geometry} 
\usepackage[hang, small,labelfont=bf,up,textfont=it,up]{caption} 
\usepackage{booktabs} 

\usepackage{lettrine} 

\usepackage{enumitem} 
\setlist[itemize]{noitemsep} 

\usepackage{abstract} 

\usepackage{titlesec} 
\renewcommand\thesection{\Roman{section}} 
\renewcommand\thesubsection{\roman{subsection}} 
\titleformat{\section}[block]{\large\scshape\centering}{\thesection.}{1em}{} 
\titleformat{\subsection}[block]{\large}{\thesubsection.}{1em}{} 

\usepackage{fancyhdr} 
\usepackage{titling} 
\usepackage{graphicx}
\usepackage{hyperref} 

\pretitle{\begin{center}\Huge\bfseries} 
\posttitle{\end{center}} 
\title{Recurrent U-net for automatic pelvic floor muscle segmentation on 3D ultrasound.} 
\author{%
\textsc{Frieda van den Noort} \\[1ex] 
\normalsize University of Twente \\ 
\normalsize \href{mailto:f.vandennoort@utwente.nl}{f.vandennoort@utwente.nl} 
\and 
\textsc{Beril Sirmacek} \\[1ex] 
\normalsize J\"{o}nk\"{o}ping University \\
\and 
\textsc{Cornelis H. Slump}\\[1ex]
\normalsize University of Twente
}
\date{\today} 
\renewcommand{\maketitlehookd}{%
\begin{abstract}
The prevalance of pelvic floor problems is high within the female population. Transperineal ultrasound (TPUS) is the main imaging modality used to investigate these problems. Automating the analysis of TPUS data will help in growing our understanding of pelvic floor related problems. In this study we present a U-net like neural network with some convolutional long short term memory (CLSTM) layers to automate the 3D segmentation of the levator ani muscle (LAM) in TPUS volumes. The CLSTM layers are added to preserve the inter-slice 3D information. We reach human level performance on this segmentation task. Therefore, we conclude that we successfully automated the segmentation of the LAM on 3D TPUS data. This paves the way towards automatic in-vivo analysis of the LAM mechanics in the context of large study populations.
\end{abstract}
}

\begin{document}

\maketitle


\section{Introduction}
\label{intro}
Even though pelvic floor related problems are common among women, both in science and society, the severeness of the issue goes largely unnoticed. Around a quarter (older than 20 years \cite{Nygaard2008}) to one-third (older than 40 years \cite{Rortveit2010}) of women experience problems, like pelvic organ prolapse, urinary and fecal incontinence, at some point in their live. Patients are often ashamed of their problems and doctors often consider these problems as collateral damage to vaginal birth \cite{Skinner2018}, a consequence of the forces put on the pelvic floor during delivery. Even though this is indeed  one of the main risk factors, the problem is more complicated. Since in most women these problems do not occur directly after delivery, but after menopause. They might even affect women who have delivered via Cesarean section or who never were pregnant \cite{Nygaard2008,Blomquist2018}. The problems occurring seem to be a result of an not yet understood interplay between different factors, such as the (damaged) pelvic floor muscles, ligaments, connective tissue, hormones and abdominal pressure \cite{Delancey2016}. Ultrasound is a convenient way to image the pelvic floor, because it is low-cost and widely available. However, transperineal ultrasound (TPUS), is complicated to interpret and research towards quantitative analysis is still in its early stages. And while 4D data is available, most investigation is limited to manual 1D or 2D measurements on specific volume frames \cite{Dietz2017,vandenNoort2019}.

Automated analysis of TPUS might be the key to better understand pelvic floor problems. The main pelvic floor muscle complex, the levator ani (LAM), appears bright on TPUS, due to the high collagen concentration in its extracellular matrix  \cite{Tuttle2014}. Automating the segmentation of this muscle in TPUS volumes may lead to a better understanding of its (dys)function and its relationship with pelvic floor problems \cite{VandenNoort2018}. TPUS data is easily collected in clinical practice, so we can analyze a statistically significant amount of data. Furthermore, since the data is 4D data (ultrasound volumes as time series), we can obtain functional information from the muscle. 

Most of the 2D image segmentation problems could be solved successfully \cite{Litjens2017}, after the successful re-introduction of convolutional neural networks (CNN) into the mainstream image processing community by \cite{Krizhevsky2012}. Especially U-net showed excellent results in most 2D segmentation applications \cite{Ronneberger2015}, since it simultaneously encodes the image data into relevant image features while keeping spatial information to get accurate high resolution segmentations. However, 3D segmentation is more complicated since training full resolution deep CNNs quickly runs into GPU memory problems. Furthermore, ultrasound data itself presents additional challenges due to the scattering within most tissues it has a noisy appearance. This makes it hard to discriminate borders between different structures and presents high variability between different operators and machines of different manufacturers \cite{Liu2019}, even for the LAM on TPUS \cite{VandenNoort2018}. 

In this paper we propose a new neural network architecture to for  segmentation of the LAM on 3D TPUS data. Our network architecture consists of 2D U-net \cite{Ronneberger2015} with some convolution layers replaced by a convolutional long short term memory cell (CLSTM) \cite{Shi2015} to transfer 3D information between the slices. The background section of this paper will provide an overview of the literature of TPUS ultrasound segmentation and other relevant literature on 3D image segmentation. In the materials and methods section we will discuss our dataset and our proposed network design for segmenting this dataset. In the results section we will present the training results compared to traditional 2D U-net on this data, to show that the network learns inter slice information due to the CLSTM. In the final section of this paper we will discuss our results to end with a final conclusion.    

\section{Background}

In this background section we will first discuss literature on segmentation on female TPUS data. Secondly, we will discuss the approaches used for segmentation of 3D image data. We will discuss in more detail the use of recurrent neural networks (RNN): such as long sort therm memory cells (LSTM) and CLSTM, for 3D image segmentation, since these methods are closely related to our approach. In the last subsection we will focus on the challenges of the segmentation of the LAM on TPUS data, in order to know what should be focused on in our network design.   

\subsection{Segmentation of transperineal ultrasound}
Currently, few publications exist on automatic segmentation of structures on female TPUS, when the writing of this paper is finalized. Segmentation of the urogenital hiatus and the LAM on a specific clinically relevant slice (2D) was successfully achieved with CNNs in two studies \cite{Bonmati2018,vandenNoort2019}. We have performed successful automatic 3D segmentation of the LAM using an active appearance model (AAM) \cite{VandenNoort2018}. However, AAMs are prone to large texture and shape variations within the data and therefore do not generalize well. CNNs became the state-of-the-art for segmentation, however AAM still can be combined with deep learning algorithms, where the AAM segmentation serves as initialization for the CNN \cite{Cheng2016}.  The most recent paper on TPUS segmentation is by \cite{Williams2019}, segmenting the urethra on TPUS, using a patched-based 3D CNN approach. 

\subsection{3D segmentation}
The problem of 3D segmentation is more complex then 2D segmentation, as briefly mentioned in the introduction. Deep network architectures like U-net \cite{Ronneberger2015} or ResNet \cite{He2016} proved to have superior segmentation results. However, often the GPU memory is the limiting factor in the network design and the required memory grows exponentially if the same architecture for 2D segmentation would be expanded to 3D segmentation. Slice by slice 2D segmentation may be a solution for 3D segmentation. However, successful 3D segmentation often requires more contextual information than is available on a single slice. 

Although this is not a review article we will provide a brief overview of some of the solutions that are suggested to address these problems. One line of approach are the 2.5D segmentation methods, which is the umbrella term for methods that aim to maximize the contextual information in slice-by-slice or voxel classification segmentation. In some of these approaches additional (usually 2) parallel \cite{Alkadi2019} or orthogonal \cite{Roth2014} slices are provided. Others use pre-processing to map surrounding information into smaller samples \cite{Angermann2019}. Finally we mention the idea to train multiple 2D segmentation networks for the three orthogonal image orientations and combined these segmentations by a small 3D CNN \cite{Xue2020}. 

Closely related to the 2.5D approaches are the patch-based methods. Here the 3D volume is divided into small volume patches. These patches are used for classification of the center voxel \cite{Milletari2017} or voxels \cite{Chen2018} or for segmentation of the complete patch \cite{Li2017}. The results of the voxel classification or segmentation of the patches are then combined into a segmentation of the complete volume, usually complemented by some post processing algorithms. 

Another approach is multi-scale processing of the data, which also has several ways it can be approached. For example, this can be achieved in a sequential order where the data is down-sampled to a lower image size, yielding a low resolution segmentation which is up-sampled to the original image size. Both this coarse segmentation and the original image are then cropped into smaller (but full resolution) image patches, which are fed to a secondary network in order to refine segmentation \cite{Roth2018}. Another possibility is a parallel method, where a certain region of interest is fed to the network at full resolution, together with down-sampled images of the surrounding area. This can be done at multiple resolutions and area sizes. The various image-scales are used as input to separate branches of the network and combined only at the final layers of the network \cite{Kamnitsas2017,Li2018}.

\subsection{Segmentation using RNNs}

Finally, RNNs can be used in the training for 3D segmentation. RNNs are networks that are trained to process sequential data, like for example language. There are many types of RNN implementation, they all focus on remembering relevant information in a sequence of data in order to properly preform the task they are designed to do. RNNs can therefore also be used for segmentation, since evaluating the image for example pixel-by-pixel or slice-by-slice also creates a sequential data stream. 

One of the most used RNN cell is the LSTM \cite{Hochreiter1997}. Besides remembering relevant information it is also trained to actively forget irrelevant information. There are several variants of the LSTM cell. Most LSTMs consists of an input gate $i_t$, memory cell $c_t$, forget gate $f_t$, output gate $o_t$ and hidden state $h_t$. Input data $x_t$ can flow through a LSTM cell as follows \cite{Gers2002,Graves2013};
\begin{equation}
i_t = \sigma (W_{xi}x_t +W_{hi}h_{t-1} +W_{ci} \circ c_{t-1} + b_i)
\end{equation}
\begin{equation}
f_t = \sigma (W_{xf}x_t +W_{hf}h_{t-1} +W_{cf} \circ c_{t-1} + b_f) 
\end{equation}
\begin{equation}
c_t = f_t \circ c_{t-1} + i_t \circ \tanh(W_{xc}x_t +W_{hc}h_{t-1} + b_c) 
\end{equation}
\begin{equation}
o_t = \sigma (W_{xo}x_t +W_{ho}h_{t-1} +W_{co} \circ c_t + b_o) 
\end{equation}
\begin{equation}
h_t = o_t \circ \tanh(c_t).
\end{equation}
The various weight matrices $W$ and bias vectors $b$ in these equations are the trainable parameters of this LSTM cell. The activation functions $\sigma$ and $\tanh$ are the sigmoid and tangent hyperbolic function, $\circ$ denotes element wise multiplication. The hidden and cell state are passed from one step in the input sequence to the next. 

Before CNN became state-of-the-art in segmentation, LSTMs were also used for segmentation. A pixel is segmented  by feeding the contextual pixels in a specified order to a LSTM-RNN \cite{Graves2007}. This is however not the most efficient way of training in the context of image processing, especially when used for 3D segmentation. \cite{Stollenga2015} proposed a different training scheme incorporating convolutions in the LSTM. Another approach to incorporate recurrency in CNNs is by sharing feature maps between slices \cite{Liang2015, Alom2018}.  Almost at the same time, \cite{Shi2015} introduced the CLSTM cell, which differs slightly from the LSTM cell;
\begin{equation}
i_t = \sigma (W_{xi}\ast X_t +W_{hi}\ast H_{t-1} +W_{ci} \circ C_{t-1} + b_i)
\end{equation} 
\begin{equation}
f_t = \sigma (W_{xf}\ast X_t +W_{hf}\ast H_{t-1} +W_{cf} \circ C_{t-1} + b_f)
\end{equation}
\begin{equation}
C_t = f_t \circ C_{t-1} + i_t \circ \tanh(W_{xc}\ast X_t +W_{hc}\ast H_{t-1} + b_c) 
\end{equation}
\begin{equation}
o_t = \sigma (W_{xo}\ast X_t +W_{ho}\ast H_{t-1} +W_{co} \circ C_t + b_o) 
\end{equation}
\begin{equation}
H_t = o_t \circ \tanh(C_t).
\end{equation}
$X_t$, $H_t$ and $C_t$ are 3D tensors, but by introducing the convolution ($\ast$) the weight matrices can be much smaller than needed in a conventional LSTM cell. This makes it very suitable for the analysis of sequential image data, like movies or slices of volume images, since deeper networks can be build.

Several approaches incorporating CLSTMs in CNNs have thus been suggested for movie \cite{Akilan2019,Arbelle2019,Salvador2017} or volume segmentation \cite{Chen2016a,Tseng2017,Li2019,Cai2017}. 
They are often based on auto-encoder segmentation network designs, where a CLSTM is incorporated within the network or applied after the network as a type of post processing. Some studies use a U-net type network as their base network \cite{Arbelle2019, Salvador2017,Chen2016a}. For example, \cite{Arbelle2019} incorporate the CLSTM in the U-net layers, showing good segmentation result in microscopy movie segmentations. These studies prove that incorporating of CLSTMs in CNNs benefits segmentation results in sequential image data.

\begin{figure}[h!]
\centering
\includegraphics[width=\linewidth]{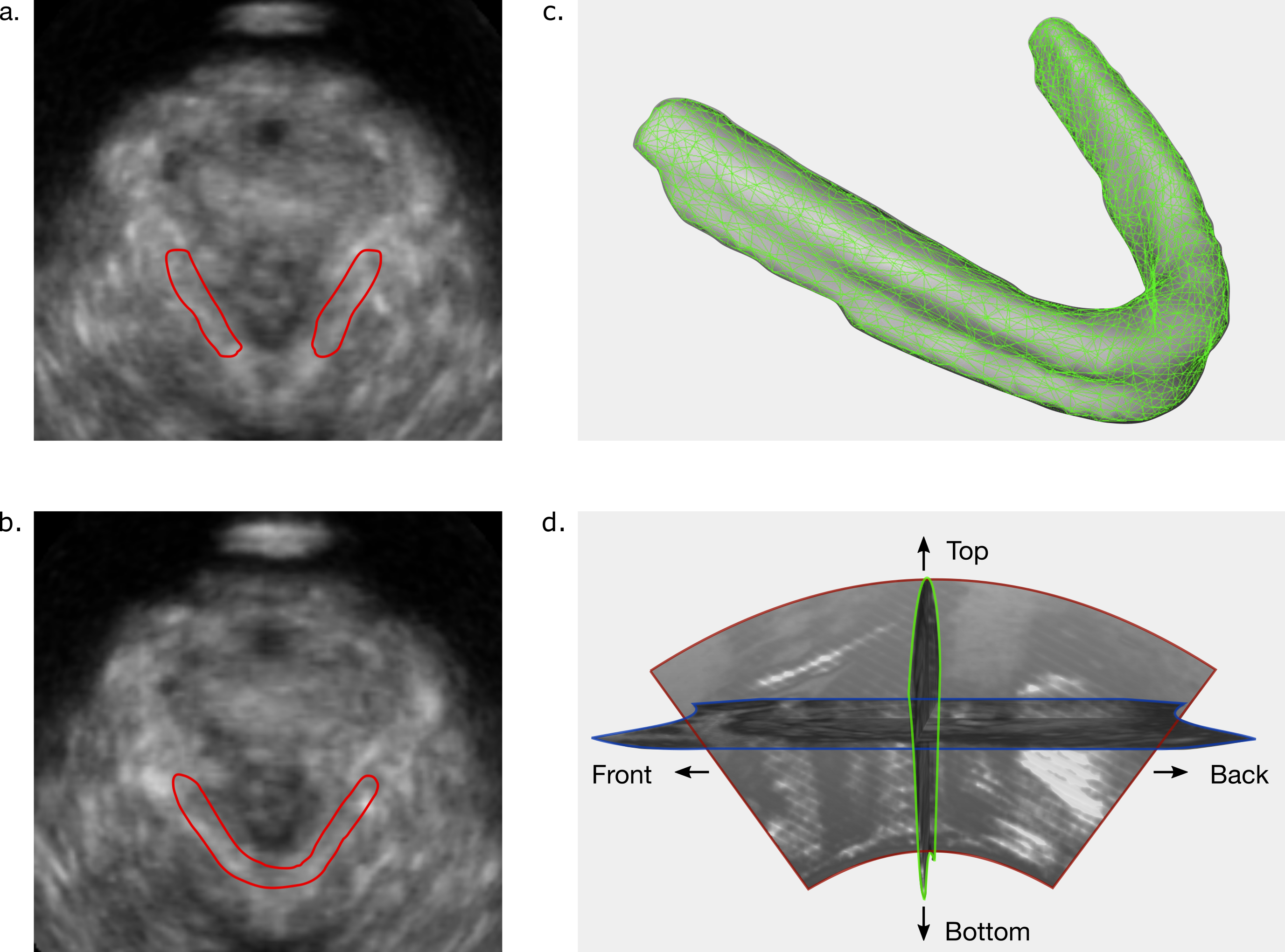}
\caption{a. and b. show examples of axial slices looking similar but requiring different segmentation (red) based on 3D information obtained for the sagital and coronal view, highlighting the necessity of 3D contextual information for the segmentation of TPUS data. c. shows an example of a segmented LAM volume. d. visualizes three orthogonal slices form the ultrasound volume (green: coronal, blue: axial, red: sagital) and their orientation in the body. This examplifies the curved apex shape of the TPUS data.}
\label{fig1}
\end{figure}

\subsection{TPUS segmentation challenges}
\label{challenges}
3D segmentation of TPUS is a challenging task due to the noisy appearance of the data. The boundaries of the LAM  are hard to distinguish due to the noisy appearance of ultrasound.  Manual segmentation of this data is already challenging and requires the possibility to rotate the volume to slices that are not in one of the three standard planes and single slice segmentation requires anatomical references from the other views \cite{VandenNoort2018}. To exemplify this, we provide two axial slices (Figure 1 a. and b.) which look similar, but have different segmentations due to information from the coronal and sagital view. Therefore, we should incorporate large context information in our network design. Furthermore, we should have a deep network design: Low-level features are important for correct segmentation, since high-level features, such as local intensity differences, are not reliable due to the noisy boundaries. 

Figure 1 c. shows the volume of a complete 3D muscle segmentation, from which can be observed that the LAM is a thin elongated structure ($\sim$1 cm in diameter and $\sim$12 cm long \cite{vandenNoort2019}). This makes segmentation of the LAM challenging for both manual and automatic segmentation, since a few voxels mismatch already can account for a relatively large volume shift. Furthermore, the class imbalance is large with only ~0.5$\%$ of the voxels being muscle. The network training should be optimized to compensate for this. The automatic segmentation is considered successful if the segmentation accuracy is on par with manual segmentation accuracy between human observers \cite{VandenNoort2018}. 
\section{Materials and Methods}
\label{MaM}
\subsection{Data}
The TPUS data used in this study originates from the dataset acquired by  \cite{VanVeelen2013}. Approval for this study was obtained from the Medical Research Ethics Committee (MREC) of the UMC Utrecht and all women signed informed consent forms. The data was acquired with a GE Voluson 730 Expert system (GE Healthcare, Zipf, Austria) equipped with an RAB 4-8MHz curved array volume transducer. The following measurement settings were used: volume angles 85$^\circ$ longitudinal and 70$^\circ$ transverse, a temporal resolution 3 Hz, gain 15, power 100, harmonics mid, contrast 8, grey map 4, persistence 8, enhance 3, depth 6 cm, fixed time gain compensation (all buttons aligned in the center). During the acquisition participants were in supine position and the probe was placed on the perineum. Before examination they were asked to empty their bladder. 

We have randomly selected TPUS data of 100 women from this dataset. Those women were around 12 weeks' pregnant and had not delivered before. We only included data on which the pelvic floor muscle and the pelvic bone were fully captured, since this is important for a successful manual segmentation \cite{VandenNoort2018}. In order to use the ultrasound data it had to be converted to DICOM manually, using 4DView (GE Healthcare, Zipf, Austria), meaning that the volume size had to be manually selected. This resulted in a variable output volume size of 215-230 by 235-250 by 140-150 slices in resp. the \textit{x, y} and \textit{z} direction. 

We imported the data in MeVisLab (MeVis Medical Solutions, Bremen, Germany \cite{Ritter2011}) to manually segment the LAM on a single volume-frame at which the LAM was at rest (the movies captured muscle contraction and stretching as well). The data was further processed in Python after segmentation, the data size was made uniform by enlarging the volumes with zeros of a 256$\times$256$\times$154 volume, since this was a convenient volume size for axial training. We used a volume size of 256$\times$256$\times$128 for training with coronal and sagital slices. This results in a slightly cropped TPUS volume, however the ultrasound image shape is a curved apex so the top and bottom never contain muscle information (Figure 1 d.). The spatial resolution differs due to different depth settings during acquisition, voxels are cubic and $\approx$ 0.5 mm along each dimension. 

The dataset was randomly split into a training- (85 volumes), validation- (5 volumes) and testset (10 volumes).

\begin{figure}[h!]
\centering
\includegraphics[width=\linewidth]{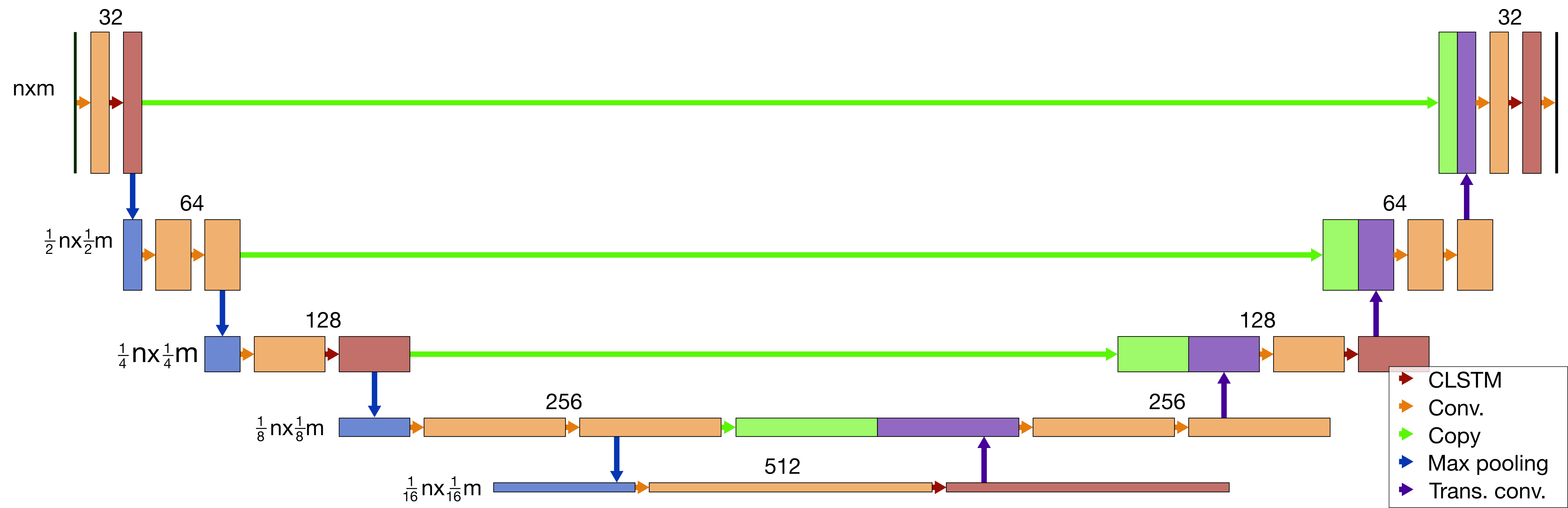}
\caption{Design of the recurrent-U-net. The input-size of a single image slice is n by m. The arrows denote an operation like applied to the preceding network layer. The convolution operations (orange) are performed with a kernel of 3x3 and result in the next layer with the filter depth denoted above the layer. The convolutional long short term memory cells (CLSTM) get their information from the previous image slice via their cell-state, which updates their 3x3 kernel used for the convolution operation. The copy operation duplicates a layer from one side of the network and concatenates it with another layer. The max pooling operation down-samples the layer by applying a 2x2 kernel with stride 2 and only selects the maximum value to the next layer. The last operation is the transposed convolution (Trans. conv.) by which the data is up-sampled with a 3x3 kernel and stride 2. }
\label{fig2}
\end{figure} 

\subsection{Network design}
As mentioned in the introduction, U-net \cite{Ronneberger2015} is the state of the art network for 2D segmentation. It gives a solution to the trade-off between spatial reduction, to get more general image features, and keeping spatial information, in order to get a spatially refined segmentation. The spatial down-sampling is obtained by max-pooling in the encoding part of the network and spatial up-sampling is obtained by a transposed convolution in the decoding part of the network. In order to use the spatial information from encoding layers into the decoding layers, the output from encoding layers was concatenated with the input to decoding layers, with corresponding spatial resolution. A similar network design for 3D data, with a reasonable input size would quickly run into memory issues. 

To circumvent this problem and solve the challenges mentioned in the last subsection of the background, we choose a 2D U-net design for our network. In order to also extract the 3D spatial information necessary for 3D segmentation we replace for every odd spatial dimension the second convolution layer with a CLSTM layer, see Figure 2. This creates a pathway to transfer 3D information from on slice to the next, in both encoding and decoding branch on different spatial feature levels. To make the training updates generalize over multiple slices, i.e. use a batch of several slices for each training step, we had to make a trade-off in GPU memory usage. Therefore, we did not incorporate a CSLTM layer on each feature level and only used half of the filters normally utilized in U-net. The network design used in this paper is referred to as the recurrent U-net (RU-net).

\subsection{Implementation}
Three networks were trained the investigate the influence of spatial orientation on the network. We also summed the results of the network, since it contains information of all three networks this is expected to yield a slightly better result compared to the individual networks. Furthermore, to show the effect of adding CLSTM layers to a U-net we also train a standard U-net on the data in all three orientations, to compare the results. 

The networks were trained using the Python language (Tensorflow 1.15) on a computer equipped with two Titan Xp graphics cards. The Dice similarity index (DSI) loss-function \cite{Milletari2016} and the Adam-optimizer \cite{Kingma2015} ($\beta_1 = 0.9$, $\beta_2=0.999$) were used for training. The loss function was masked so that it was only calculated on voxels part of the TPUS image.  The Rectified Linear Unit (ReLU) \cite{Nair2010} was used as activation function in the network and a sigmoid function was used for the output layer. During training, a dropout with rate 0.3 was applied after each convolution layer. The network trained for 100 epochs with a learning rate of 10$^{-5}$, and was saved when the highest validation DSI was achieved. The training data was randomized for each epoch. Each volume was processed in subsequent batches of 14 slices, when training with axial slices, and 16 slices, when training with coronal and sagital slices. To minimize the impact of the class imbalance and to speed-up the convergence of the training, the batches containing muscle labels were trained twice during a training step, until a validation DSI of 0.5 was reached. Since U-net does not transfer any knowledge for one slice to the next one we also randomized the slices for every training step of the U-net. 

\subsection{Testing}

To test the results of the network we applied a threshold following the sigmoid activation function of the output layer: Voxels with value $\geq$ 0.5 were assigned to the muscle label, the others to the background. To investigate the success of the segmentations we calculated the DSI, mean absolute distance (MAD) and Hausdorff distance (HDD) between the network segmentations and the manual labels. The data was cleaned by image dilation and erosion and small objects were removed. 
\begin{figure}[h!]
\centering
\includegraphics[width=\linewidth]{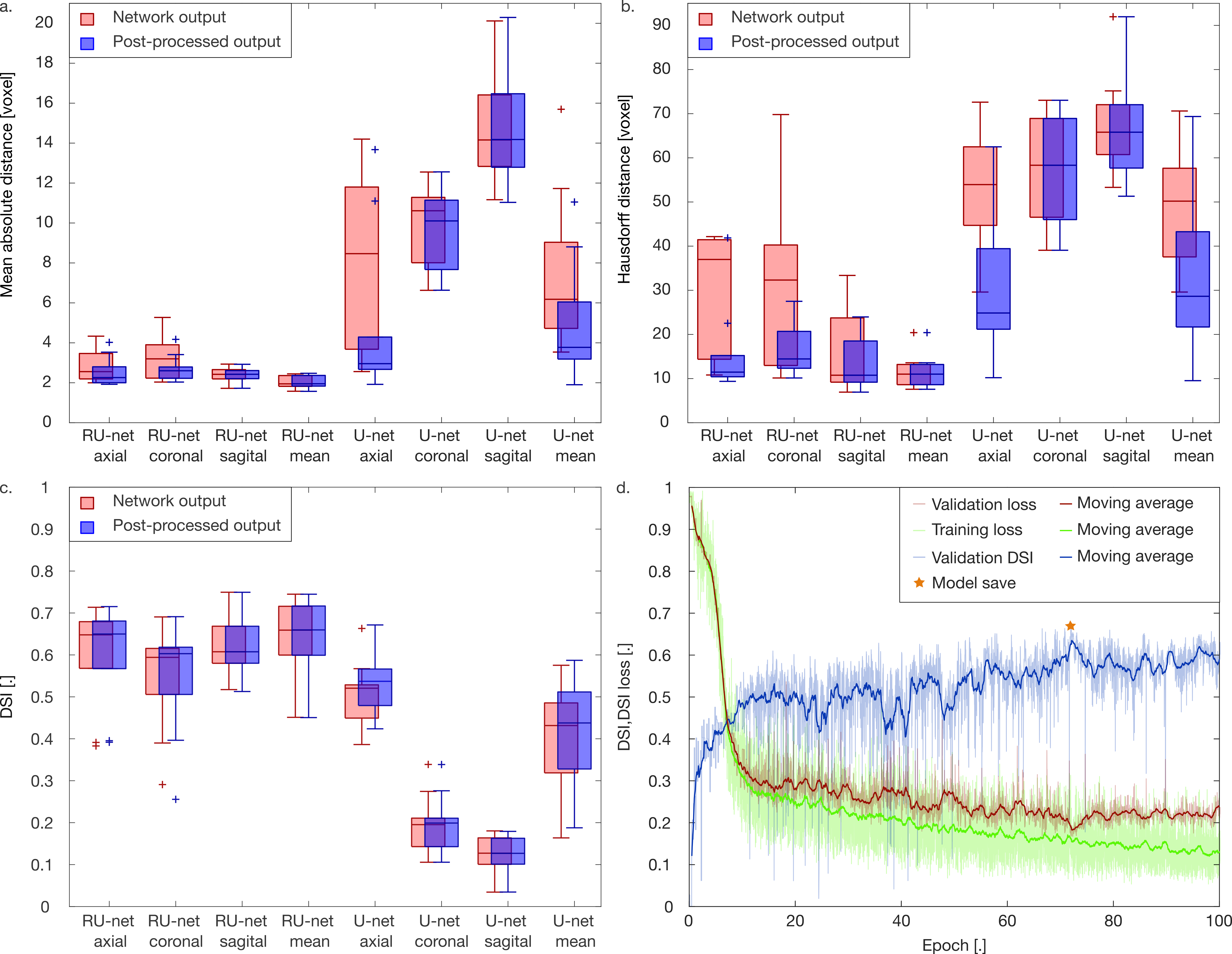}
\caption{a.-c. show the boxplots of mean absolute distance, Hausdorff distance and Dice simularity index (DSI) between the manual and network segmentations for both recurrent-U-net (RU-net) and U-net. Boxes with internal lines represent median and inter-quartile range (IQR), whiskers are range excluding outliers (+) larger than 1.5 IQR from upper and lower quartile. Both networks were trained for the three different spatial orientations (which means they are either trained with axial, coronal or sagital slices) and the mean of the output of these networks is calculated. The values between manual segmentation and the original output of the network are shown in red, whereas the results after post-processing are displayed in blue. d. shows the training and validation DSI loss and the DSI of the validation data. A moving average with a kernel of 1 epoch is plotted over the data and used to better visualize trends in the data.}
\label{fig3}
\end{figure}

\section{Results}
\label{Results}

The networks for different orientations were trained and Figure 3 a.-c. show box-plots with the MAD, HDD and DSI of the test results for RU-net and U-net for the 3 different slice orientations and their mean, with and without post-processing. Since the spatial resolution differs between volumes the distances are measured in voxels instead of mm. The results reflect the fact that incorporating CLSTM in U-net improves the segmentation results significantly. Figure 3 d. shows the training progress, DSI and DSI loss, of the network trained on coronal slices. Since the network is trained with slice batches and not with different data batches, the training and validation losses fluctuate for each training step. A moving average filter of 1 epoch is applied, to better visualize the training and validation trends. Due to this fluctuation, the validation DSI and loss were checked after every training step, therefore the total training time of 1 network was $\sim$ 50 hours. Deploying the network to a single test sample takes $\sim$ 3 seconds. 

Segmenting with RU-net on sagital slices preforms best as individual network, which is surprising since, U-net performs worst on sagital slices. Still, U-net performs relatively well on axial slices this is as expected, since manual segmentation is also performed on slightly tilted axial slices, with some feedback of the sagital and coronal view. 

Individual RU-nets benefit from post processing, since there are small blobs segmented in the image which are not part of the LAM. This benefits the MAD and HDD, but barely influences the DSI, as can be seen in Figure 3 a.-c. However, combining the 3 orientation makes post-processing almost irrelevant. This makes the results on average better than one individual RU-net, although the results are very close to the sagital RU-net. 

Visual inspection of the average RU-net results show that they mostly produce smooth surfaces that resemble the manual segmentation. There are 2 exceptions  worth mentioning here, both having a DSI $\le$ 0.6. The worst result correctly segments the LAM 'legs' but misses the `sling', which seems to be due to the fact that the `sling' is almost outside image view. The other result correctly segments the LAM volume, however the result is not smooth. No clear explanation can be found when looking to the original image: It does not appear significantly different from other TPUS images.

\section{Discussion}
\label{Discuss}
In this study we show that the LAM can be successfully segmented on TPUS volumes, by upgrading 2D U-net with CLSTM layers to transfer contextual information between slices. The results are better than inter observer - and comparable to intra observer results between manual observers \cite{VandenNoort2018}. Since the training data was segmented manually by one observer, this means that the network was able to learn the segmentation 'style' of this observer. The results were also comparable to the results achieved for the urethra segmentation on TPUS \cite{Williams2019}.

Our network solves the challenges stated in last subsection of the background section. With the U-net design of the network, we were able to capture both low and high level image features of the slices. Replacing some convolutional layers in U-net with CLSTM layers allowed for transfer of these features between slices to capture the 3D image context. The training process was straightforward, it turned out that no data sampling was required to compensate for the class imbalance: Training twice with slice batches containing muscle data did slightly speed up the training, but did not improve the final results. Finally, the results were on par with human segmentation performance. 

The quick deployment time of the network makes it suitable to apply in clinical practice. From a clinical perspective it is also interesting to observe that sagital RU-net achieved better results than coronal and axial RU-net. Since the sagital view seems to contain information relevant for 3D segmentation, investigate this view more closely for clues that could aid manual segmentation. 

One can debate on were to put the CLSTM layers in U-net (or simular networks). We decided on putting these in both the encoding and decoding part of the network at different spatial resolution levels. Others decide to put it only on the encoding or decoding part of the network \cite{Arbelle2019,Salvador2017}. 
Since the results of our design are already on par with intra observer results, which should be the upper limit, we decided not to exhaustively test all possible options. The weaker segmentations are more likely explained by a lack of training data: Since most LAM are not located next to the edge there is not much training data on this.

We tested retraining the network from one slice orientation to the next. In theory it is to be expected that most feature layers should be shared between the networks. However, this did achieve better training results compared to a network trained with random initialization. The training loss decreased quicker, however due to the stochastic nature of the training, there seemed to be less chance on a good validation result, before over-fitting occurred. We adopted the default parameters for the Adam-optimizer, since changing them did not improve the results. Successful training of the network was sensitive to the learning rate: small changes would cause the network either to stop training after a few epochs, or to have very poor learning. 

The slices can be fed to the network in two directions, for example the axial slices can be fed from top to bottom or from bottom to top (Figure 1 d.). The results did not change for sagital and coronal slices. However the axial slices had to be fed from top to bottom. It seemed during training that the curved bottom of the TPUS data and the shape of the LAM sling were closely related in shape. Therefore, the network learned emphasizing the bottom curvature and corners, but, since this does not contain relevant information, training was severely slowed down. This problem can be prevented by not feeding the bottom part of the TPUS data, instead we decided to change the order of the slices from top to bottom.

\section{Conclusions}
\label{Conclusions}
In this study we incorporated CLSTM layers into a U-net to segment the LAM on 3D TPUS images. We showed that incorporating these layers improves the segmentation success of U-net so that it is on par with human manual segmentation performance. We can therefore conclude that we successfully automated the  segmentation of the LAM on 3D TPUS data. This paves the way towards automatic in-vivo analysis of the LAM mechanics in the context of large study populations.

\section*{Acknowledgements}
\label{Ack}
We gratefully acknowledge the support of NVIDIA Corporation with the donation of the Titan Xp GPU used for this research

\section*{Conflicts of interest}
The authors declare no conflict of interest. 


\bibliographystyle{ieeetr}
\bibliography{citations}

\begin{thebibliography}{10}

\bibitem{Nygaard2008}
I.~Nygaard, M.~D. Barber, K.~L. Burgio, K.~Kenton, S.~Meikle, J.~Schaffer,
  C.~Spino, W.~E. Whitehead, J.~Wu, and D.~J. Brody, ``{Prevalence of
  Symptomatic Pelvic Floor Disorders in US Women},'' {\em Jama}, vol.~300,
  no.~11, p.~1311, 2008.

\bibitem{Rortveit2010}
G.~Rortveit, L.~L. Subak, D.~H. Thom, J.~M. Creasman, E.~Vittinghoff, S.~K.
  {Van den Eeden}, and J.~S. Brown, ``{Urinary incontinence, fecal incontinence
  and pelvic organ prolapse in a population-based, racially diverse cohort:
  Prevalence and risk factors},'' {\em Female Pelvic Medicine and
  Reconstructive Surgery}, vol.~16, no.~5, pp.~278--283, 2010.

\bibitem{Skinner2018}
E.~M. Skinner, B.~Barnett, and H.~P. Dietz, ``{Psychological consequences of
  pelvic floor trauma following vaginal birth: a qualitative study from two
  Australian tertiary maternity units},'' {\em Archives of Women's Mental
  Health}, vol.~21, pp.~341--351, jun 2018.

\bibitem{Blomquist2018}
J.~L. Blomquist, A.~Mu{\~{n}}oz, M.~Carroll, and V.~L. Handa, ``{Association of
  Delivery Mode with Pelvic Floor Disorders after Childbirth},'' {\em JAMA -
  Journal of the American Medical Association}, vol.~320, pp.~2438--2447, dec
  2018.

\bibitem{Delancey2016}
J.~DeLancey, ``{Chapter Two: Pelvic Floor Anatomy and Pathology},'' in {\em
  Biomechanics of the Female Pelvic Floor} (L.~Hoyte and M.~Damaser, eds.),
  pp.~13--51, Amsterdam: Elsevier Inc., 1rst~ed., 2016.

\bibitem{Dietz2017}
H.~P. Dietz, ``{Pelvic Floor Ultrasound: A Review},'' {\em Clinical Obstetrics
  and Gynecology}, vol.~60, no.~1, pp.~58--81, 2017.

\bibitem{vandenNoort2019}
F.~van~den Noort, C.~H. van~der Vaart, A.~T.~M. Grob, M.~K. van~de Waarsenburg,
  C.~H. Slump, and M.~van Stralen, ``{Deep learning enables automatic
  quantitative assessment of puborectalis muscle and urogenital hiatus in plane
  of minimal hiatal dimensions},'' {\em Ultrasound in Obstetrics {\&}
  Gynecology}, vol.~54, pp.~270--275, aug 2019.

\bibitem{Tuttle2014}
L.~J. Tuttle, O.~T. Nguyen, M.~S. Cook, M.~Alperin, S.~B. Shah, S.~R. Ward, and
  R.~L. Lieber, ``{Architectural design of the pelvic floor is consistent with
  muscle functional subspecialization},'' {\em International Urogynecology
  Journal and Pelvic Floor Dysfunction}, vol.~25, no.~2, pp.~205--212, 2014.

\bibitem{VandenNoort2018}
F.~van~den Noort, A.~T.~M. Grob, C.~H. Slump, C.~H. van~der Vaart, and M.~van
  Stralen, ``{Automatic segmentation of puborectalis muscle on
  three‐dimensional transperineal ultrasound},'' {\em Ultrasound in
  Obstetrics {\&} Gynecology}, vol.~52, pp.~97--102, jul 2018.

\bibitem{Litjens2017}
G.~Litjens, T.~Kooi, B.~E. Bejnordi, A.~Arindra, A.~Setio, F.~Ciompi,
  M.~Ghafoorian, J.~A. W.~M. {Van Der Laak}, B.~{Van Ginneken}, and C.~I.
  S{\'{a}}nchez, ``{A Survey on Deep Learning in Medical Image Analysis},''
  {\em Medical image analysis}, vol.~42, pp.~60--88, 2017.

\bibitem{Krizhevsky2012}
A.~Krizhevsky, I.~Sutskever, and G.~E. Hinton, ``{ImageNet Classification with
  Deep Convolutional Neural Networks},'' {\em Proceedings of the 25th
  International Conference on Neural Information Processing Systems}, vol.~1,
  pp.~1097--1105, 2012.

\bibitem{Ronneberger2015}
O.~Ronneberger, P.~Fischer, and T.~Brox, ``{U-Net: Convolutional Networks for
  Biomedical Image Segmentation},'' {\em International Conference on Medical
  image computing and computer-assisted intervention}, pp.~234--241, 2015.

\bibitem{Liu2019}
S.~Liu, Y.~Wang, X.~Yang, B.~Lei, L.~Liu, S.~X. Li, D.~Ni, and T.~Wang, ``{Deep
  Learning in Medical Ultrasound Analysis: A Review},'' {\em Engineering},
  vol.~5, pp.~261--275, apr 2019.

\bibitem{Shi2015}
X.~Shi, Z.~Chen, H.~Wang, D.-Y. Yeung, W.-K. Wong, and W.-C. Woo,
  ``Convolutional lstm network: A machine learning approach for precipitation
  nowcasting,'' in {\em Proceedings of the 28th International Conference on
  Neural Information Processing Systems - Volume 1}, NIPS’15, (Cambridge, MA,
  USA), p.~802–810, MIT Press, 2015.

\bibitem{Bonmati2018}
E.~Bonmati, Y.~Hu, N.~Sindhwani, H.~P. Dietz, J.~D'hooge, D.~Barratt,
  J.~Deprest, and T.~Vercauteren, ``{Automatic segmentation method of pelvic
  floor levator hiatus in ultrasound using a self-normalizing neural
  network},'' {\em Journal of Medical Imaging}, vol.~5, no.~2, p.~021206, 2018.

\bibitem{Cheng2016}
R.~Cheng, H.~R. Roth, L.~Lu, S.~Wang, B.~Turkbey, W.~Gandler, E.~S. McCreedy,
  H.~K. Agarwal, P.~Choyke, R.~M. Summers, and M.~J. McAuliffe, ``{Active
  appearance model and deep learning for more accurate prostate segmentation on
  MRI},'' {\em Medical Imaging 2016: Image Processing}, vol.~9784, p.~97842I,
  2016.

\bibitem{Williams2019}
H.~{Williams}, L.~{Cattani}, W.~{Li}, M.~{Tabassian}, T.~{Vercauteren},
  J.~{Deprest}, and J.~{D’hooge}, ``3{D} convolutional neural network for
  segmentation of the urethra in volumetric ultrasound of the pelvic floor,''
  in {\em 2019 IEEE International Ultrasonics Symposium (IUS)}, pp.~1473--1476,
  2019.

\bibitem{He2016}
K.~He, X.~Zhang, S.~Ren, and J.~Sun, ``{Deep residual learning for image
  recognition},'' in {\em Proceedings of the IEEE Computer Society Conference
  on Computer Vision and Pattern Recognition}, vol.~2016-December,
  pp.~770--778, IEEE Computer Society, dec 2016.

\bibitem{Alkadi2019}
R.~Alkadi, A.~El-Baz, F.~Taher, and N.~Werghi, ``{A 2.5D deep learning-based
  approach for prostate cancer detection on T2-weighted magnetic resonance
  imaging},'' in {\em Proceedings of the European Conference on Computer Vision
  (ECCV),}, vol.~11132 LNCS, pp.~734--739, 2019.

\bibitem{Roth2014}
H.~R. Roth, L.~Lu, A.~Seff, K.~M. Cherry, J.~Hoffman, S.~Wang, J.~Liu,
  E.~Turkbey, and R.~M. Summers, ``{A new 2.5D representation for lymph node
  detection using random sets of deep convolutional neural network
  observations},'' in {\em Medical Image Computing and Computer-Assisted
  Intervention -- MICCAI 2014}, vol.~8673 LNCS, pp.~520--527, Springer Verlag,
  2014.

\bibitem{Angermann2019}
C.~Angermann and M.~Haltmeier, ``{Random 2.5D U-net for Fully 3D
  Segmentation},'' in {\em Machine Learning and Medical Engineering for
  Cardiovascular Health and Intravascular Imaging and Computer Assisted
  Stenting.}, pp.~158--166, Springer, 2019.

\bibitem{Xue2020}
Y.~Xue, F.~G. Farhat, O.~Boukrina, A.~M. Barrett, J.~R. Binder, U.~W. Roshan,
  and W.~W. Graves, ``{A multi-path 2.5 dimensional convolutional neural
  network system for segmenting stroke lesions in brain MRI images},'' {\em
  NeuroImage: Clinical}, vol.~25, no.~April 2019, p.~102118, 2020.

\bibitem{Milletari2017}
F.~Milletari, S.-A. Ahmadi, C.~Kroll, A.~Plate, V.~Rozanski, J.~Maiostre,
  J.~Levin, O.~Dietrich, B.~Ertl-Wagner, K.~B{\"{o}}tzel, and N.~Navab,
  ``{Hough-CNN: Deep learning for segmentation of deep brain regions in MRI and
  ultrasound},'' {\em Computer Vision and Image Understanding}, vol.~164,
  no.~0, pp.~92--102, 2017.

\bibitem{Chen2018}
L.~Chen, Y.~Wu, A.~M. DSouza, A.~Z. Abidin, A.~Wism{\"{u}}ller, and C.~Xu,
  ``{MRI tumor segmentation with densely connected 3D CNN},'' in {\em Medical
  Imaging 2018: Image Processing}, p.~105741F, SPIE-Intl Soc Optical Eng, jan
  2018.

\bibitem{Li2017}
W.~Li, G.~Wang, L.~Fidon, S.~Ourselin, M.~J. Cardoso, and T.~Vercauteren, ``{On
  the compactness, efficiency, and representation of 3D convolutional networks:
  Brain parcellation as a pretext task},'' in {\em International conference on
  information processing in medical imaging.}, vol.~10265 LNCS, pp.~348--360,
  Springer Verlag, jul 2017.

\bibitem{Roth2018}
H.~R. Roth, C.~Shen, H.~Oda, T.~Sugino, M.~Oda, Y.~Hayashi, K.~Misawa, and
  K.~Mori, ``{A Multi-scale Pyramid of 3D Fully Convolutional Networks for
  Abdominal Multi-organ Segmentation},'' in {\em International conference on
  medical image computing and computer-assisted intervention}, vol.~11073 LNCS,
  pp.~417--425, Springer Verlag, 2018.

\bibitem{Kamnitsas2017}
K.~Kamnitsas, C.~Ledig, V.~F. Newcombe, J.~P. Simpson, A.~D. Kane, D.~K. Menon,
  D.~Rueckert, and B.~Glocker, ``{Efficient multi-scale 3D CNN with fully
  connected CRF for accurate brain lesion segmentation},'' {\em Medical Image
  Analysis}, vol.~36, pp.~61--78, feb 2017.

\bibitem{Li2018}
X.~Li, Q.~Dou, H.~Chen, C.~W. Fu, X.~Qi, D.~L. Belav{\'{y}}, G.~Armbrecht,
  D.~Felsenberg, G.~Zheng, and P.~A. Heng, ``{3D multi-scale FCN with random
  modality voxel dropout learning for Intervertebral Disc Localization and
  Segmentation from Multi-modality MR Images},'' {\em Medical Image Analysis},
  vol.~45, pp.~41--54, 2018.

\bibitem{Hochreiter1997}
S.~Hochreiter and J.~Schmidhuber, ``{Long Short-Term Memory},'' {\em Neural
  Computation}, vol.~9, no.~8, pp.~1735--1780, 1997.

\bibitem{Gers2002}
F.~A. Gers, N.~N. Schraudolph, and J.~Schmidhuber, ``{Learning precise timing
  with LSTM recurrent networks},'' {\em Journal of Machine Learning Research},
  vol.~3, no.~1, pp.~115--143, 2002.

\bibitem{Graves2013}
A.~Graves, ``{Generating Sequences With Recurrent Neural Networks},'' {\em
  arXiv preprint arXiv:1308.0850}, aug 2013.

\bibitem{Graves2007}
A.~Graves, S.~Fernandez, and J.~Schmidhuber, ``{Multi-Dimensional Recurrent
  Neural Networks},'' in {\em International conference on artificial neural
  networks}, pp.~549--558, Springer, 2007.

\bibitem{Stollenga2015}
M.~F. Stollenga, W.~Byeon, M.~Liwicki, and J.~Schmidhuber, ``{Parallel
  Multi-Dimensional LSTM, With Application to Fast Biomedical Volumetric Image
  Segmentation},'' {\em arXiv preprint arXiv:1506.07452}, jun 2015.

\bibitem{Liang2015}
M.~Liang and X.~Hu, ``{Recurrent convolutional neural network for object
  recognition},'' in {\em Proceedings of the IEEE conference on computer vision
  and pattern recognition}, pp.~3367--3375, 2015.

\bibitem{Alom2018}
M.~Z. Alom, M.~Hasan, C.~Yakopcic, T.~M. Taha, and V.~K. Asari, ``{Recurrent
  Residual Convolutional Neural Network based on U-Net (R2U-Net) for Medical
  Image Segmentation},'' {\em arXiv preprint arXiv:1802.06955}, feb 2018.

\bibitem{Akilan2019}
T.~Akilan, Q.~J. Wu, A.~Safaei, J.~Huo, and Y.~Yang, ``{A 3D CNN-LSTM-Based
  image-to-image foreground segmentation},'' {\em IEEE Transactions on
  Intelligent Transportation Systems}, vol.~21, pp.~959--971, mar 2019.

\bibitem{Arbelle2019}
A.~Arbelle and T.~R. Raviv, ``{Microscopy cell segmentation via convolutional
  LSTM networks},'' in {\em 2019 IEEE 16th International Symposium on
  Biomedical Imaging (ISBI 2019)}, pp.~1008--1012, IEEE, 2019.

\bibitem{Salvador2017}
A.~Salvador, M.~Bellver, V.~Campos, M.~Baradad, F.~Marques, J.~Torres, and
  X.~Giro-I-Nieto, ``{Recurrent Neural Networks for Semantic Instance
  Segmentation},'' {\em arXiv preprint arXiv:1712.00617}, 2017.

\bibitem{Chen2016a}
J.~Chen, L.~Yang, Y.~Zhang, M.~Alber, and D.~Z. Chen, ``{Combining Fully
  Convolutional and Recurrent Neural Networks for 3D Biomedical Image
  Segmentation},'' {\em arXiv preprint arXiv:1609.01006}, sep 2016.

\bibitem{Tseng2017}
K.~L. Tseng, Y.~L. Lin, W.~Hsu, and C.~Y. Huang, ``{Joint sequence learning and
  cross-modality convolution for 3D biomedical segmentation},'' in {\em
  Proceedings of the IEEE conference on Computer Vision and Pattern
  Recognition}, pp.~6393--6400, 2017.

\bibitem{Li2019}
H.~Li, J.~Li, X.~Lin, and X.~Qian, ``{MDS-Net: A Model-Driven Stack-Based Fully
  Convolutional Network for Pancreas Segmentation},'' {\em arXiv preprint
  arXiv:1903.00832}, mar 2019.

\bibitem{Cai2017}
J.~Cai, L.~Lu, Y.~Xie, F.~Xing, and L.~Yang, ``{Improving Deep Pancreas
  Segmentation in CT and MRI Images via Recurrent Neural Contextual Learning
  and Direct Loss Function},'' {\em arXiv preprint arXiv:1707.04912}, jul 2017.

\bibitem{VanVeelen2013}
G.~A. van Veelen, K.~J. Schweitzer, and C.~H. van~der Vaart, ``{Reliability of
  pelvic floor measurements on three- and four-dimensional ultrasound during
  and after first pregnancy: Implications for training},'' {\em Ultrasound in
  Obstetrics and Gynecology}, vol.~42, pp.~590--595, nov 2013.

\bibitem{Ritter2011}
B.~F. Ritter, T.~Boskamp, A.~Homeyer, H.~Laue, M.~Schwier, F.~Link, and H.-O.
  Peitgen, ``{A Visual Approach},'' {\em IEEE Pulse}, no.~November 2011,
  pp.~60--70, 2011.

\bibitem{Milletari2016}
F.~Milletari, N.~Navab, and S.~A. Ahmadi, ``{V-Net: Fully convolutional neural
  networks for volumetric medical image segmentation},'' in {\em Proceedings -
  2016 4th International Conference on 3D Vision, 3DV 2016}, pp.~565--571,
  IEEE, oct 2016.

\bibitem{Kingma2015}
D.~P. Kingma and J.~L. Ba, ``{Adam: A method for stochastic optimization},'' in
  {\em 3rd International Conference on Learning Representations, ICLR 2015 -
  Conference Track Proceedings}, International Conference on Learning
  Representations, ICLR, dec 2015.

\bibitem{Nair2010}
V.~Nair and G.~E. Hinton, ``{Rectified Linear Units Improve Restricted
  Boltzmann Machines},'' in {\em Proceedings of the 27th International
  Conference on International Conference on Machine Learning}, pp.~807--814,
  omnipress~ed., 2010.

\end{thebibliography}


\end{document}